\documentclass{aa}

\usepackage{longtable,lscape,graphicx,float,txfonts,natbib}

\begin{document}

\title{The RMS Survey: Mid-Infrared Observations of Candidate Massive YSOs in the Southern Hemisphere\thanks{Based on observations collected at the European Southern Observatory, La Silla, Chile (ESO Programmes 70.C-0069(A), 73.C-0314(A), 74.C-0267 and 077.C-0687(A))}}

\author{J.C.~Mottram\inst{1}~\thanks{E-mail:jcm@ast.leeds.ac.uk}
  \and
  ~M.G.~Hoare\inst{1}
  \and
  ~S.L.~Lumsden\inst{1}
  \and
  ~R.D.~Oudmaijer\inst{1}
  \and
  ~J.S.~Urquhart\inst{1}
  \and
  ~T.L.~Sheret\inst{1}
  \and
  ~A.J.~Clarke\inst{1}
  \and
  ~J.~Allsopp\inst{2}
}

\institute{School of Physics and Astronomy, University of Leeds, Leeds, LS2 9JT, UK
  \and
  Astrophysics Research Institute, Liverpool John Moores University, Twelve Quays House, Egerton Wharf, Birkenhead, CH41 1LD, UK}

\date{Received }

\abstract
   {The Red MSX Source (RMS) survey is an ongoing effort to return a large, well-selected sample of massive young stellar objects (MYSOs) within our Galaxy. 2000 candidates have been colour-selected from the Mid-course Space Experiment (MSX) point source catalogue (PSC). A series of ground-based follow-up observations are being undertaken in order to remove contaminant objects (ultra-compact HII (UCHII) regions, planetary nebulae (PN), evolved stars), and to begin characterising these MYSOs.}
   {As a part of these follow-up observations, high resolution ($\sim$1$^{\prime}$$^{\prime}$) mid-IR imaging aids the identification of contaminant objects which are resolved (UCHII regions, PN) as opposed to those which are unresolved (YSOs, evolved stars) as well as identifying YSOs near UCHII regions and other multiple sources.}
   {We present 10.4$\mu$m imaging observations for 346 candidate MYSOs in the RMS survey in the Southern Hemisphere, primarily outside the region covered by the GLIMPSE Spitzer Legacy Survey. These were obtained using TIMMI2 on the ESO 3.6m telescope in La Silla, Chile. Our photometric accuracy is of order 0.05Jy, and our astrometric accuracy is 0.8$^{\prime}$$^{\prime}$, which is an improvement over the nominal 2$^{\prime}$$^{\prime}$ accuracy of the MSX PSC. }
   {Point sources are detected in 64$\%$ of our observations, which are expected to be either YSOs or evolved stars. 24$\%$ contain only sources of extended emission, which are likely to be either UCHII regions or, in a few cases, PN. This is confirmed by comparison with radio continuum observations. We find that, as expected for a dusty HII region, the strength of 10.4$\mu$m and radio continuum emission is related. The remaining targets (12$\%$) result in non-detections. While for 63$\%$ of our targets we detect only one mid-infrared source, 25$\%$ show multiple sources. In these cases, our observations will allow the apportioning of the flux from larger beam measurements between the different sources. Within these multiple source targets, we find some point sources on or near UCHII regions. Our improved astrometric information will allow more accurate targeting of spectroscopy, which will be used to identify unresolved sources in cases where it is not clear whether they are YSOs or evolved stars.}
   {}


\maketitle

\section{Introduction}

Massive stars (M$\geq$8M$_{\odot}$) form within dense molecular clouds of dust and gas, but evolve so rapidly that the central star cannot be observed directly before it has reached the main sequence. As a result, the formation mechanism and early evolution of massive stars is still shrouded in mystery, with little consensus in the field as to a single formation theory, as opposed to that for low mass star formation. 

A tentative observationally based evolutionary sequence is emerging, starting with dense cores in the molecular cloud and progressing through the hot molecular core and massive young stellar object (MYSO) phases \citep{b1}. Eventually, the star begins to ionise the material around it and an ultra-compact HII (UCHII) region forms, which then expands into an increasingly large HII region before the central star finally becomes visible in the optical as an OB star.

We define an MYSO to be an embedded near or mid-infrared (IR) source that is luminous enough to be a young O or B-type star, but has not yet begun to ionise the surrounding molecular gas to form an HII region. The MYSO phase is of particular interest, because this is the phase in which the star may have started core hydrogen burning, but major accretion is presumably still ongoing. MYSOs are therefore a key stage for learning about how massive stars form, and probing the various formation theories currently proposed \citep[][]{b9,b8,b7,b31}.

However, those MYSOs which have been widely observed were mainly discovered serendipitously, and are often rather nearby. Surveys with IRAS data \citep[e.g.][]{b5, b10, b11} have only considered bright, unconfused IRAS sources. As such their samples tend to avoid the dense parts of the galactic plane, due to confusion in the large IRAS beam (45$^{\prime}$$^{\prime}$ $\times$ 240$^{\prime}$$^{\prime}$ at 12$\mu$m), but this is where the majority of MYSOs should lie. As such, the previously identified MYSOs are unlikely to be representative of the class as a whole.

The mid-infrared Galactic Plane survey with the Mid-course Space Experiment (MSX) by \citet{b6} has provided better resolution data ($\sim$18$^{\prime}$$^{\prime}$ in all bands), allowing for most of these problems to be overcome. \citet{b4} developed colour selection criteria for MYSOs in the MSX point source catalogue (PSC), using the PSC fluxes. 2 Micron All-Sky Survey (2MASS) PSC fluxes \citep{b22} were used to help eliminate blue sources. These colour-cuts ($F_{8}<F_{14}<F_{21}$,~$F_{21}/F_{8}>$2,~$F_{8}/F_{K}>$5 and $F_{K}/F_{J}>$2), along with inspection to remove sources which are actually extended in the MSX observations, have returned a sample of $\sim$2000 candidate MYSOs. These are contaminated with both young and old dusty objects which have similar near and mid-IR colours (e.g. evolved stars, planetary nebulae (PN), HII regions). The Red MSX Source (RMS) Survey aims to remove the contaminants from this MYSO candidate list via a series of ground based follow up observations at radio, mm and IR wavelengths \citep{b12}. It will also provide data on the many new UCHII regions and evolved objects detected as a by-product of the survey process.

At mid-IR wavelengths, the emission around MYSOs comes primarily from the warm (e.g. $\sim$300K at 10$\mu$m) dust close to, and heated directly by, the young star. Observations of MYSOs at $K$(2.2$\mu$m) \citep[e.g. ][]{b36,b35} show significant infrared excess from hot dust emission, suggesting that dust is present right up to the sublimation radius. The spherically symmetric models of dust emission around a 34M$_{\odot}$, 2.5$\times$10$^{5}$L$_{\odot}$ O6 type star by \citet{b2} were origionally intended for UCHII regions, but do not include Lyman $\alpha$ heating in an ionised zone and so are actually more applicable to MYSOs. Their model 1 shows that the full-width at half maximum (fwhm) of intensity at 12$\mu$m (their Fig. 2a) is 0.005pc (900AU), which at their distance of 2.6kpc corresponds to 0.36$^{\prime}$$^{\prime}$. This source lies at the near end of our distance distribution and this fwhm is well below our derived limit for classification as a point source of 1.6$^{\prime}$$^{\prime}$ (see $\S$3). The luminosity of the modelled source is also at the high end for our sample and since the emission size should scale approximately as L$^{\frac{1}{2}}$ if it follows the size of T$_{sub}$, the majority of our sources should be smaller than this. This basic situation will not change substantially in a disk-like geometry; most of the mid-IR emission will still come from the dense material close to the star, e.g. the models of \citet[][]{b39} for a 8.4M$_{\odot}$ star give a size of $\sim$200AU at 12$\mu$m. However, the dust in the walls of the cavity can be heated to $\sim$300K at further distances giving weak extended emission. While there are a few rare examples of resolved dust emission at $\sim$10$\mu$m around MYSOs \citep[e.g. G35.2N ][]{b17}, the majority of observations show only unresolved emission with the current generation of facilities \citep[e.g. ][]{b18,b19,b20} and we therefore classify YSOs as point sources in the mid-IR. In the case of G35.2N we are probably viewing the system side on with respect to the disk, so the central source is too extincted to be visible. Were we to detect something similar we would expect the source to have quite a compact symmetric bipolar or monopolar appearance without strong radio continuum emission associated with it.

For an UCHII region, the models of \citet[][their Fig. 7]{b3} suggest that 300K emission comes from the entire ionised region due to Lyman $\alpha$ heating. The mid-IR and radio continuum emission will therefore have similar a size and morphology for a given UCHII regions, and so will appear extended at 1$^{\prime}$$^{\prime}$ resolution at 1kpc, as will PN. Extended mid-IR objects can therefore be removed from the MYSO candidate sample. While radio continuum imaging can detect ionised regions, and may help identify unresolved UCHII regions or PN, it cannot detect the MYSOs which may lie near or superimposed upon HII regions. Proto-PN, low-mass stars between the post-AGB and PN phases, still have a low enough T$_{eff}$ that they produce little radio continuum emission and can appear point-like at mid-IR wavelengths \citep{b28}. Evolved stars will also appear point-like, and so both will require identification using other information. In theory reddend RGB stars could be destinguished from YSOs using JHK colour-colour diagrams, but since sources selected by our colour selection criteria are either non-detections or have upper limit fluxes at $J$ this is not possible in practise. In general, evolved stars usually appear isolated, rather than in clustered star formation regions and have little molecular CO, so both these facts can be used to differentiate them from younger sources. Near-IR spectroscopy can also be used.

High-resolution mid-IR imaging allows the combined flux of larger beam observations to be apportioned among the different sources if more than one object is present within the beam (e.g. MSX). This will improve the accuracy of any information derived from these data and allow the removal of MYSO candidates from the list which are point-like but whose mid and far-IR fluxes in larger beam observations are actually dominated by extended emission from a nearby HII region. In addition, the improved astrometric information will be useful for subsequent near-IR spectroscopy as the MSX coordinates are not good enough to identify the near-IR counterpart in crowded regions.

Here we present mid-infrared observations undertaken in the southern hemisphere as part of the RMS survey. In $\S$2 we report the details of the observational process. In $\S$3 we present our results, along with discussion of the astrometry ($\S$3.1), photometry ($\S$3.2), and a comparison with radio continuum data ($\S$3.3). In $\S$4 we reach our conclusions.

\section{Observations}

Imaging observations were undertaken at 10.4$\mu$m (Filter FWHM=1.02$\mu$m) with the Thermal Infrared Multi-mode Instrument 2 \citep[TIMMI2;][]{b27} on the European Southern Observatory 3.6m telescope at La Silla, Chile over 4 runs, totalling 12 nights between 2003 and 2006. The 10.4$\mu$m filter was chosen as it does not include either PAH emission features which could lead to extended emission, or the [NeII] line at 12.8$\mu$m which can be present in ionised regions. A nod-chop observing mode was used, with chopping perpendicular to nodding, a usual chop/nod throw of 15$^{\prime}$$^{\prime}$ and a chop frequency of 6Hz. A pixel scale of 0.202$^{\prime}$$^{\prime}$/pix was used, resulting in a field of view of 64$^{\prime}$$^{\prime}$$\times$~48$^{\prime}$$^{\prime}$ for each individual image. Standard stars were observed at $\sim$2 hour intervals for flux calibration. For the majority of our observations, the seeing was no worse than 1.5$^{\prime}$$^{\prime}$ in the optical and close to the diffraction limit at 10$\mu$m.

The blind pointing positional accuracy of the 3.6m telescope is 5$^{\prime}$$^{\prime}$, and so on the 10 nights of observations from 2004 onwards the following procedure was used to improve the positional accuracy of our observations. First an optically visible reference star, chosen to be $<$1$^{\circ}$ away from the target and bright enough to be detected quickly at 10$\mu$m (m$_{V}$$<$12 and $F_{8}>$0.7Jy), was centred on a reference pixel. Then, the telescope was traversed by the known offset between the reference star and the target coordinates. This ensured that the reference pixel in each object image was centred at the MSX target coordinates. The reference stars were chosen from the SAO star catalogue for the 2 nights of observations in 2004, but in some cases the reference stars were too faint. Therefore, a higher flux limit and the more extensive TYCHO catalogue was used for subsequent observations. This mechanism was not used for the 2 nights in 2003, and so we must assume that the targets observed in these observations are at the MSX positions.

A total of 346 targets were observed. The majority are outside the Spitzer Galactic Legacy Infrared Mid-Plane Survey Extraordinaire \citep[GLIMPSE;][]{b21} region (l=10$^{\circ}$$-$65$^{\circ}$, 295$^{\circ}$$-$350$^{\circ}$, $\mid$b$\mid$$\leq$1$^{\circ}$). While some of our observations lie within this region, they were either observed before the GLIMPSE survey was released, or the GLIMPSE images are particularly difficult to interpret and so benefit from observations at slightly better spatial resolution which are not affected by emission from transiently heated PAHs. Integration times were calculated in order to detect down to the faintest MSX fluxes ($\sim$0.1Jy) if all that flux was concentrated in a point source, with times ranging from 126s to 720s. 

Data reduction and analysis were performed using IDL scripts, with the reduction consisting of addition of the individual image frames to form nod/chop pairs, addition of all the nod-chop pairs for one observation to form a total image and then mosaicing of the resultant images.

\section{Results}

Our results are presented as follows: first we report some general information on the observations as a whole. These are then followed by discussion of the table of data and figure of all targets available in the online version only. Data on the measured angular size of the target objects are also presented. Next a more detailed discussion of the measurement and accuracy of the astrometric ($\S$3.1) and photometric ($\S$3.2) data is provided. Finally, a comparison is made between our observations and current radio continuum data ($\S$3.3) both from \citet{b14} and data obtained as part of the wider RMS efforts \citep{b13}.

A total of 346 targets were observed, with the breakdown of detections and non-detections shown in Table 1. With our chosen nod/chop throw of 15$^{\prime}$$^{\prime}$ we are able to detect sources up to this scale at least. A non-detection could therefore be due to the surface brightness of the object detected in the 18$^{\prime}$$^{\prime}$ MSX beam being below our detection threshold. These objects are most likely either HII regions or PN, not MYSOs, and so are of less interest to our survey. Alternatively the flux at 9.7$\mu$m can be reduced by factors of 5 or more with respect to the 8$\mu$m continuum in particularly embedded YSOs due to the silicate absorption feature \citep{b29}. This could lead to a non-detection, but should not be taken as proof of the absence of an MYSO.

We differentiate between target regions showing single or multiple sources for several reasons. Firstly, evolved stars are unlikely to be near any other sources of mid-IR emission and so we would expect all targets showing multiple mid-IR emission components to be related to star formation. Secondly, in cases where multiple components exist, the emission levels observed in larger beam measurements, such as IRAS and MSX, must be apportioned among these components. This may mean that extended emission near an MYSO candidate dominates, in which case the candidate may not be luminous enough to be considered an MYSO. Finally, this allows us to examine how many of our targets show potential for clustering and, in the case of point sources on or near extended emission, potentially triggered star formation. Early signs of star formation are often seen in the dense gas surrounding UCHII regions \citep[e.g.][]{b41} with good theoretical reasons for believing that their formation has been triggered by compression due to the expansion of the HII region \citep{b33}.

\begin{table}
\centering
\caption{General Target Data.}
\begin{tabular}{@{}lllcc@{}}
\hline
\centering
&\multicolumn{2}{c}{Object Type}&Number&$\%$ of Total\\
\hline
\multicolumn{3}{l}{Detection}&303&87.6\\
\hline
&Single&&218&63.0\\
&&Point ($\leq$1.6$^{\prime}$$^{\prime}$)&152&43.9\\
&&Extended ($>$1.6$^{\prime}$$^{\prime}$)&66&19.1\\
\hline
&Multiple&&85&24.6\\
&&Point ($\leq$1.6$^{\prime}$$^{\prime}$)&41&11.9\\
&&Extended ($>$1.6$^{\prime}$$^{\prime}$)&17&4.9\\
&&Extended $\&$ Point&27&7.8\\
\hline
\multicolumn{3}{l}{Non-Detections}&43&12.4\\
\hline 
\multicolumn{5}{l}{The criterion for distinguishing between point and}\\
\multicolumn{5}{l}{extended sources is discussed in the $\S$3.}\\
\end{tabular}
\end{table}

The results of our astrometric, photometric and source size measurements for all targets are shown in Table 2 available in the online version only. We will further describe these measurements, and their general trends and implications below. Target object names are given as the galactic coordinates of the MSX point source position. In the case of multiple sources, these are numbered with increasing distance from the MSX position. In the case where no object was detected in our observations, a 3$\sigma$ upper limit flux has been provided. The offset of the MSX PSC \citep[V2.3][]{b32} target position from the TIMMI2 position (column 5) and between the 2MASS PSC counterpart and the TIMMI2 coordinates for each object (column 11), where a counterpart exists and is clear, are also provided. A search for a potential 2MASS point source counterpart to each TIMMI2 source was conducted for angular separations of up to 5$^{\prime}$$^{\prime}$, though these are only likely to be true counterparts for angular separations of $\leq$2.4$^{\prime}$$^{\prime}$. The counterpart coordinates and offset from the TIMMI2 source are given in columns 9-11. Where astrometric information was not available, the primary object is assumed to be at the MSX coordinates. Nearby radio continuum detections and non-detections, either in our own interferometric observations with the ATCA \citep{b13} at 3.6cm and 6cm at a spatial resolution of 1$^{\prime}$$^{\prime}$ or in the literature \citep{b14,b37,b38}, have been shown in column 12, with `$-$' indicating that no observation has been made for this target during either set of observations. Overall 318 of our targets were observed by \citet{b13} and 11 by \citet{b14}, one (G336.4917$-$01.4741) by \citet{b37} and one (G344.9766$-$01.9071) by \citet{b38}, leaving 15 with no radio continuum observation in either set. Of those sources with no radio observations, 12 are in the northern hemisphere where VLA observations have been obtained. In some cases, extended emission at 10.4$\mu$m does not have a radio continuum detection, probably due to the surface brightness of the radio continuum emission being below the detection limit and/or resolved out by the interferometry.

Figure 1 (available online only) shows on the left the total field of view (79$^{\prime}$$^{\prime}$~$\times$~63$^{\prime}$$^{\prime}$) of the mosaiced TIMMI2 images for all targets for which sources were detected. In cases where radio emission has been detected by \citet{b13}, their contours have been plotted on the TIMMI2 image, with preference given to the 6cm detections where both 6cm and 3.6cm observations were made (e.g. G331.0309+01.2056). The beam shape, size and position angle are shown in the lower left corner of the image. For targets with multiple (e.g. G260.6877$-$01.3930) or extended (e.g. G264.2918+01.4700) sources of emission, negatives and unphysical positives can lie near the real emission, due to the nodding and chopping required for mid-IR observations and subsequent mosaicing. These were removed by 'patching' the image using the STARLINK GAIA package PATCH tool to estimate the background levels from an annulus around the region to be patched. However, in regions with large amounts of extended emission, it was difficult to remove all the negatives, and in some cases difficult to tell what was real and what was not due to sources nodding or chopping onto other sources. In particular, it should be noted that while images and results for G291.2725$-$00.7198 are presented here, they are tentative in that this region contains a large amount of extended 10.4$\mu$m emission. Our aim is to show in this paper the full field of view without artifacts so that comparisons with other wavelengths, particularly radio continuum maps, can be made easily. Any apparent structures other than those reported in Table 2 should not be over-interpreted. All the TIMMI2 data and images presented in this paper are included in the RMS survey database\footnote{The RMS database is accessible at http://www.ast.leeds.ac.uk/RMS}. In particular, the raw and mosaiced un-patched images are available, in addition to the patched mosaiced images, for those wishing to do a more in-depth comparison with their own data. The intensity scale of all images ranges from the median to the maximum level observed. For most images the scale is linear, but for those with both bright and faint sources on the same image a square root scale has been used.

The right hand images in figure 1 show the $K$ band 2MASS image for the target using a square-root intensity scale. The MSX target positions are indicated with a '+', marking the 3$\sigma$ positional errors based on comparisons between the positions of our sources and the MSX coordinates for the whole sample (5.7$^{\prime}$$^{\prime}$, see $\S$3.1). TIMMI2 detections are marked with an 'X' indicating our 3$\sigma$ astrometric accuracy (2.4$^{\prime}$$^{\prime}$), derived from comparisons of a subsample with associated 2MASS sources (again, see $\S$3.1). For simple sources the TIMMI2 and MSX positions agree very well, but in more complex star formation regions, the MSX position can be 5-10$^{\prime}$$^{\prime}$ away from the dominant point source at 10.4$\mu$m (e.g. G260.6877$-$01.3930). It should be noted that mid and near-IR sources do not always correspond to the same object, despite being close to each other since mid-IR objects can be heavily extincted in the near-IR. One should therefore not regard the nearness of sources of emission in TIMMI2 and 2MASS as an implicit indication that they come from the same source, though this may often be the case.

Source sizes were measured using a 2-D gaussian fit to the source on the image, with the results for the major and minor axes shown in figure 2. In order to make a decision on whether an object is resolved or not, we examined the distribution of source sizes near the peak. The apparent size of an object is a combination of object physical size, seeing and the diffraction limit of the 3.6m telescope (0.7$^{\prime}$$^{\prime}$ at 10$\mu$m). To remove the latter from our decision of what the maximum unresolved size of an object is, a gaussian fit was performed to the size distribution data after a correction for the beam had been made using $\theta_{source}$=($\theta_{obs}^{2}$-$\theta_{beam}^{2}$)$^{\frac{1}{2}}$. Only the data to the left of the peak was used in the fit, since all of these objects should be unresolved. The fit was then re-convolved to fit the measured source sizes. This returned a skewed gaussian with a slightly smaller spread on the lower side ($\sigma_{1}$) than the higher side ($\sigma_{2}$), since the addition in quadrature of $\theta_{source}$ and $\theta_{beam}$=0.7$^{\prime}$$^{\prime}$ has more effect on smaller values of $\theta_{source}$. Using 3$\sigma_{2}$ above the mean from this fit to the major axis FWHM we define the maximum size of a point source as 1.6$^{\prime}$$^{\prime}$, as used in table 1. The difference between the peak of the major and minor FWHM distributions (1.23$^{\prime}$$^{\prime}$ and 1.07$^{\prime}$$^{\prime}$ respectively) is primarily due to the fact that the point spread function for TIMMI2 on the 3.6m telescope is slightly elliptical rather than spherical. This difference (0.14$^{\prime}$$^{\prime}$) is actually less than the size of one pixel (0.202$^{\prime}$$^{\prime}$).

\setcounter{table}{2}
\setcounter{figure}{1}

\begin{figure*}
\centering
\includegraphics[width=80mm, height=60mm]{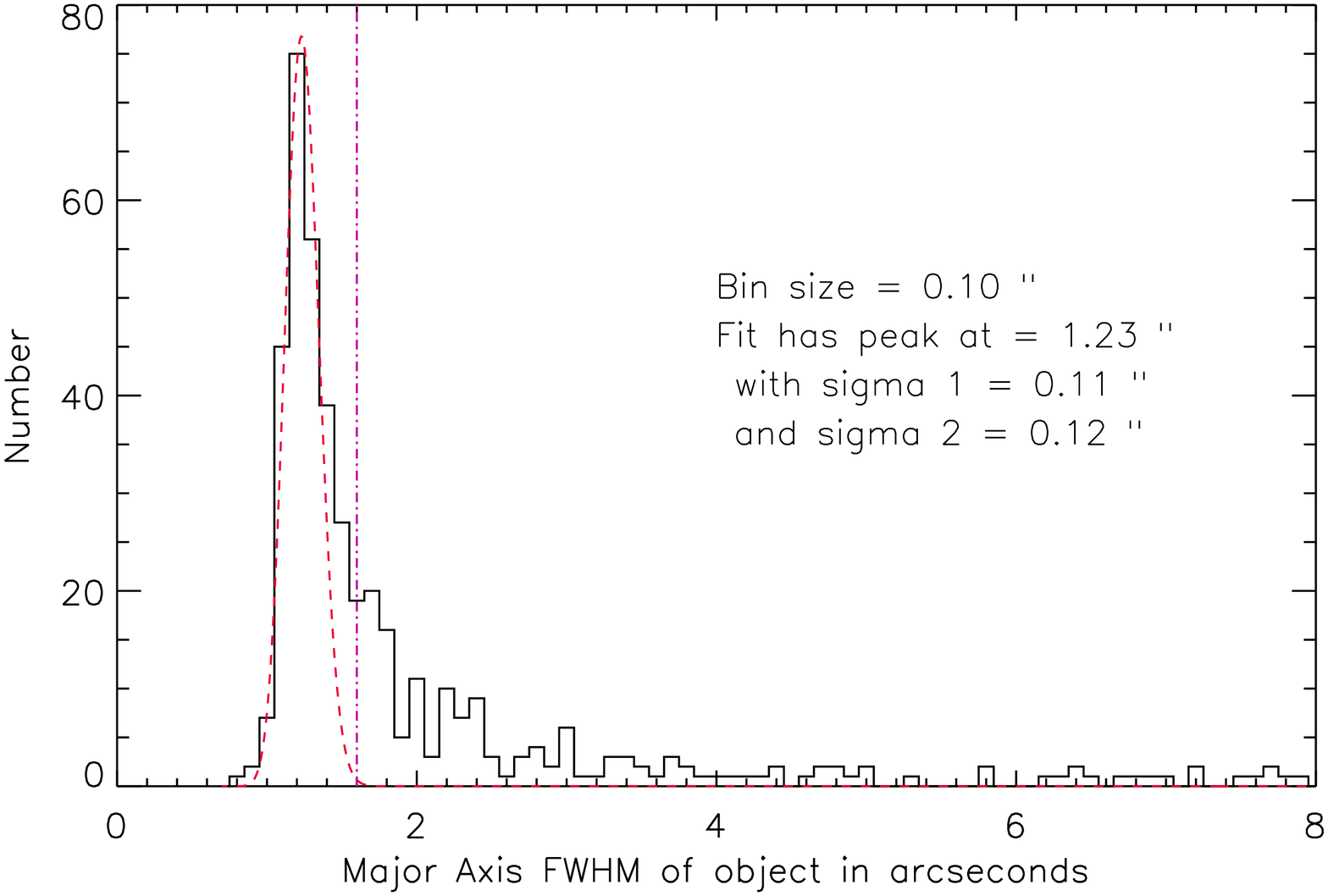}
\includegraphics[width=80mm, height=60mm]{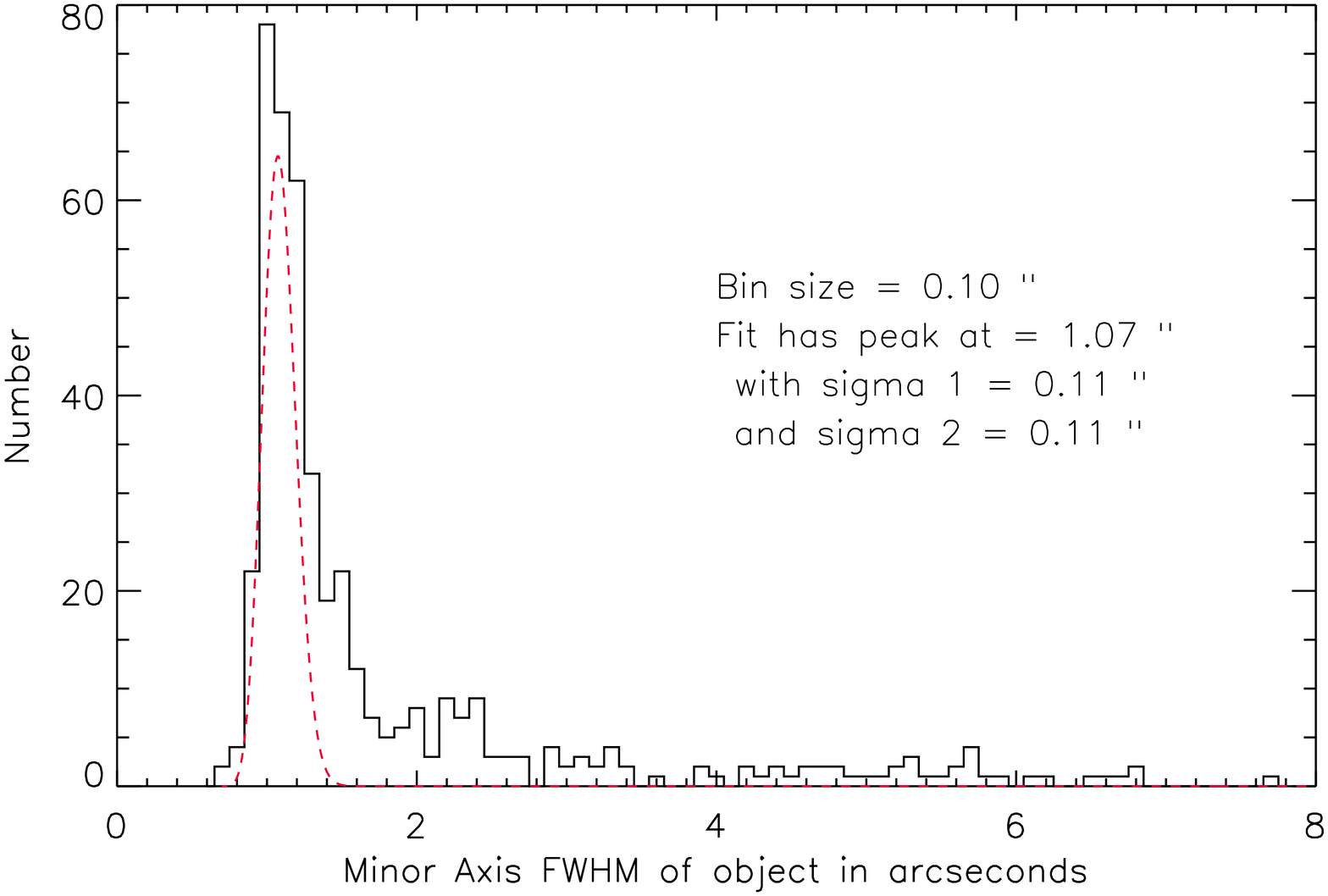}
\caption{Major and minor FWHM size of detected targets. The vertical dashed-dot line in the left hand plot shows the 1.6$^{\prime}$$^{\prime}$ boundary between point and extended sources.  The dashed lines show skewed gaussian fits to the data.}
\end{figure*}

\subsection{Astrometry}

\begin{figure*}
\centering
\includegraphics[width=80mm, height=60mm]{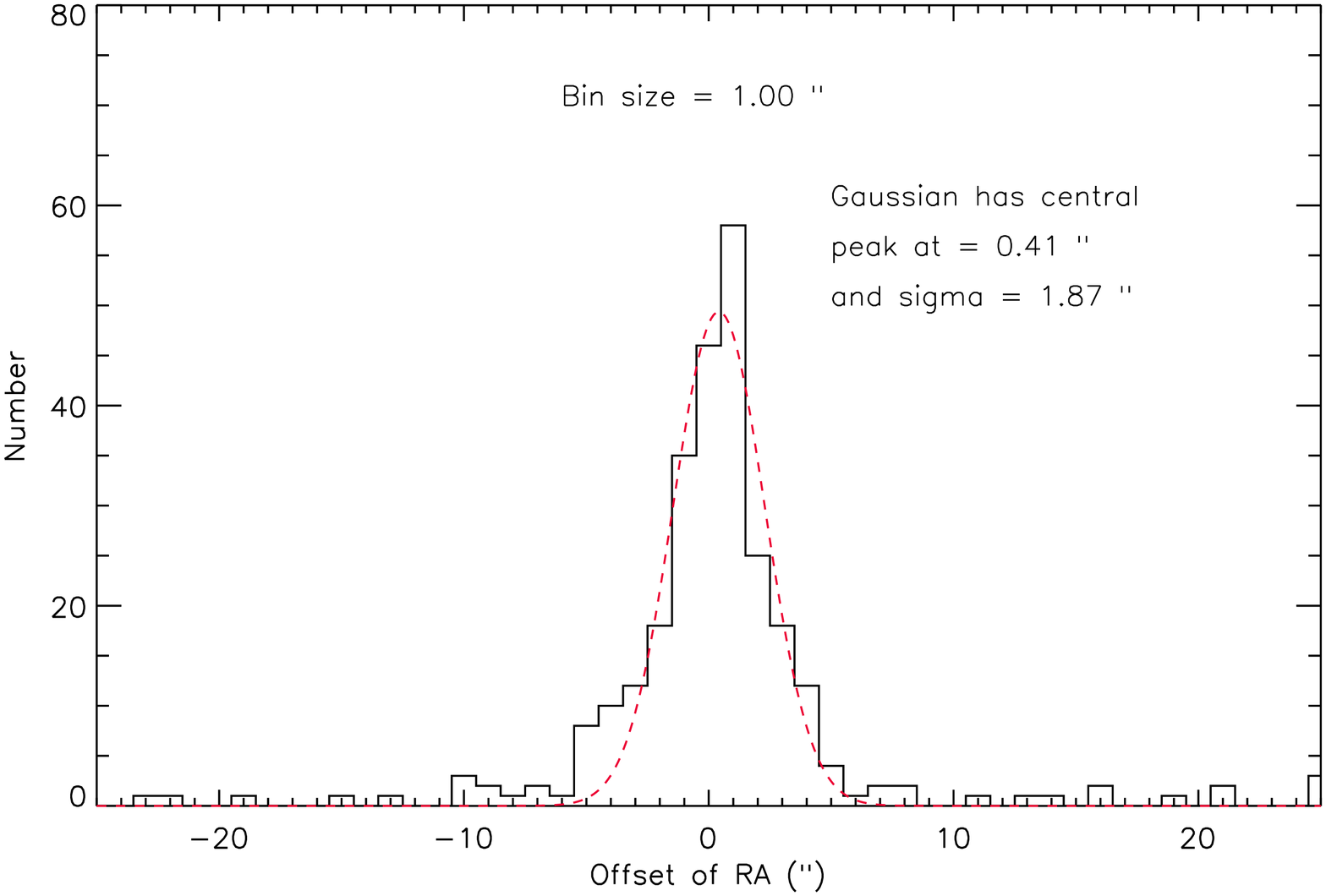}
\includegraphics[width=80mm, height=60mm]{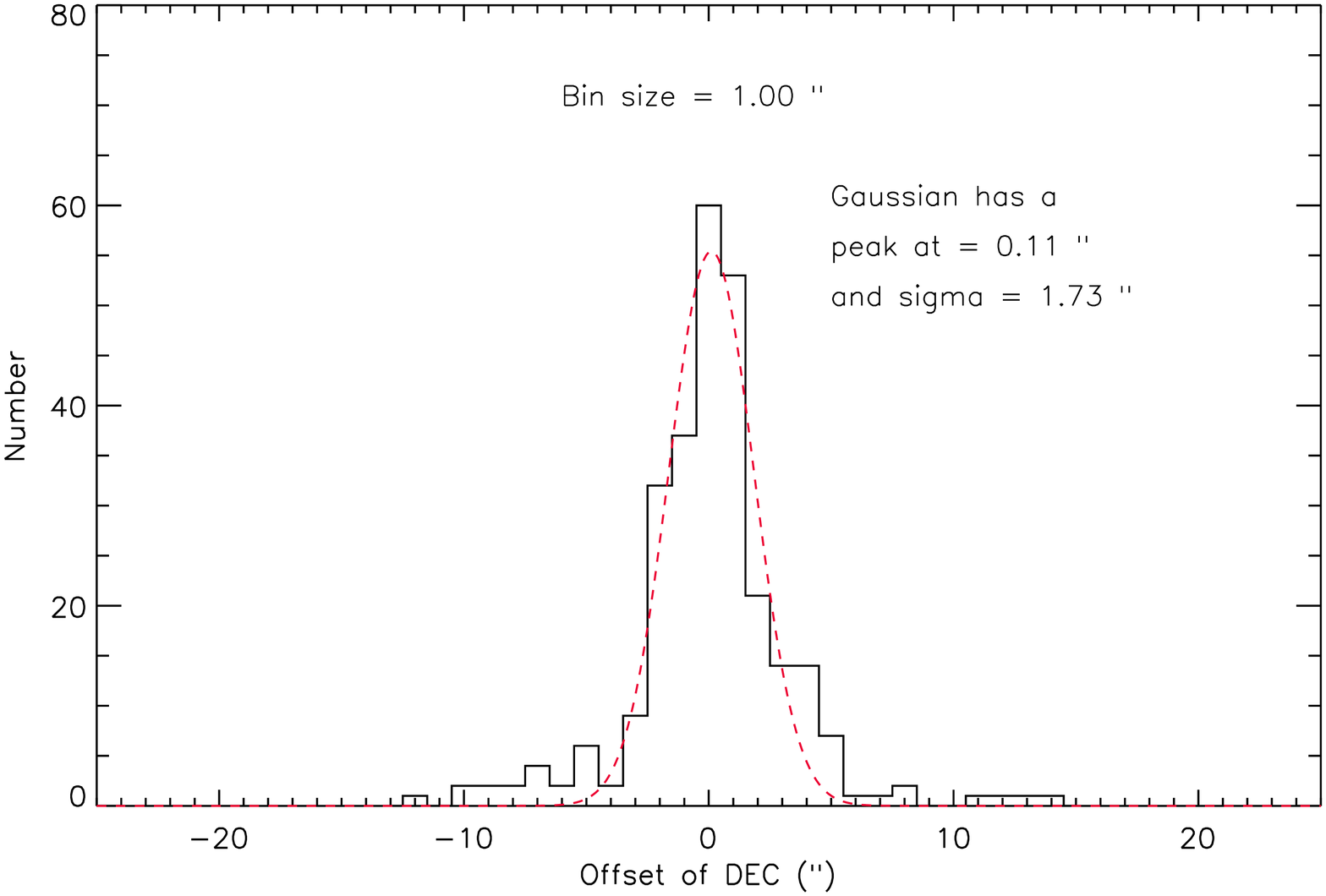}
\caption{Offset of TIMMI2 astrometry from MSX coordinates, in both Ra and Dec, for all data with astrometry. The dashed lines show gaussian fits to the data.}
\end{figure*}

Astrometry was carried out for all observations from 2004 onwards as described in $\S$2. The position of the centre of the object was obtained using a 2-dimensional gaussian fit to the source and the difference between this position and the reference pixel measured. Some of the targets with multiple sources from the 2003 observations were observed again in 2006 in order to provide accurate astrometry, which were then applied to the earlier observations, which had longer exposure times. The offsets from the MSX target coordinates of all targets with astrometry are plotted in figure 3, with the offset for the primary source being used in cases where multiple sources are detected. The offsets are similar between RA and Dec, and return an uncertainty (1$\sigma$) in the mean offset of 1.9$^{\prime}$$^{\prime}$ in RA and 1.7$^{\prime}$$^{\prime}$ in Dec (see Fig. 3). However, this assumes that the primary source in our observations corresponds to the MSX source, which is not always true. For example, in the case of the primary source for G341.1281$-$00.3466 (see Fig. 1), MSX detects one slightly extended source while we see one bright and two fainter point sources near a region of extended emission. The mean offset for both RA and Dec is not zero, but is small and well within the 2$^{\prime}$$^{\prime}$ quoted positional error of the MSX catalogue. What is more, previous comparisons with the MSX PSC have also found a non-zero total offset \citep[e.g. ][]{b4,b13}.

Let us now consider a subsample of the data for which we expect the TIMMI2 to MSX positions to be in good agreement. Sources which show only one unresolved source in our observations and are free of extended emission in both the TIMMI2 and MSX images are unlikely to have their positions skewed by diffuse emission or multiplicity. These sources are likely to be predominantly isolated evolved stars, which further reduces confusion. We find that the offsets for these objects are, on average, slightly smaller than those for all targets (see Fig. 4.), returning an astrometric accuracy of 1.4$^{\prime}$$^{\prime}$ in RA and 1.1$^{\prime}$$^{\prime}$ in Dec. However, if we compare the TIMMI2 source positions with those for the source counterpart in the more accurate (0.2$^{\prime}$$^{\prime}$) 2MASS PSC, where one exists, we get a much better correlation (0.8$^{\prime}$$^{\prime}$) than comparisons with MSX (see Fig. 4.). Since we have chosen this subsample in order to ensure that there is no confusion between counterparts in the different observations, we reach the conclusion that the offset between TIMMI2 and MSX positions is dominated by the errors in the MSX positions. On the other hand, the offset between TIMMI2 and 2MASS coordinates gives a more accurate representation of the astrometric accuracy in our TIMMI2 observations (0.8$^{\prime}$$^{\prime}$). The asymmetry seen in the offsets for this subsample between TIMMI2 and MSX are not seen when comparing between TIMMI2 and 2MASS, suggesting that this is due to systematics in the MSX positions and not those of our observations. 

\begin{figure*}
\centering
\includegraphics[width=80mm, height=60mm]{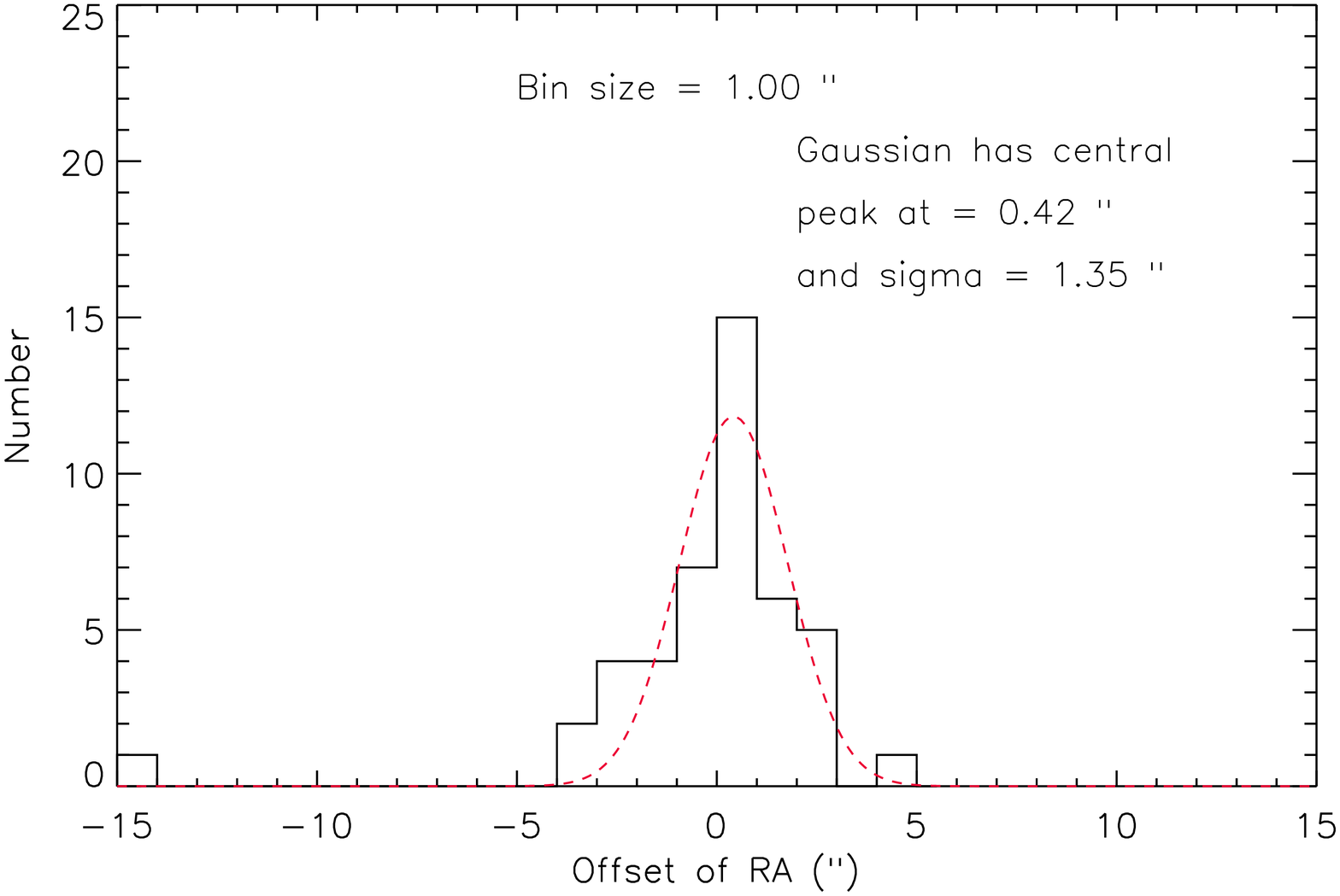}
\includegraphics[width=80mm, height=60mm]{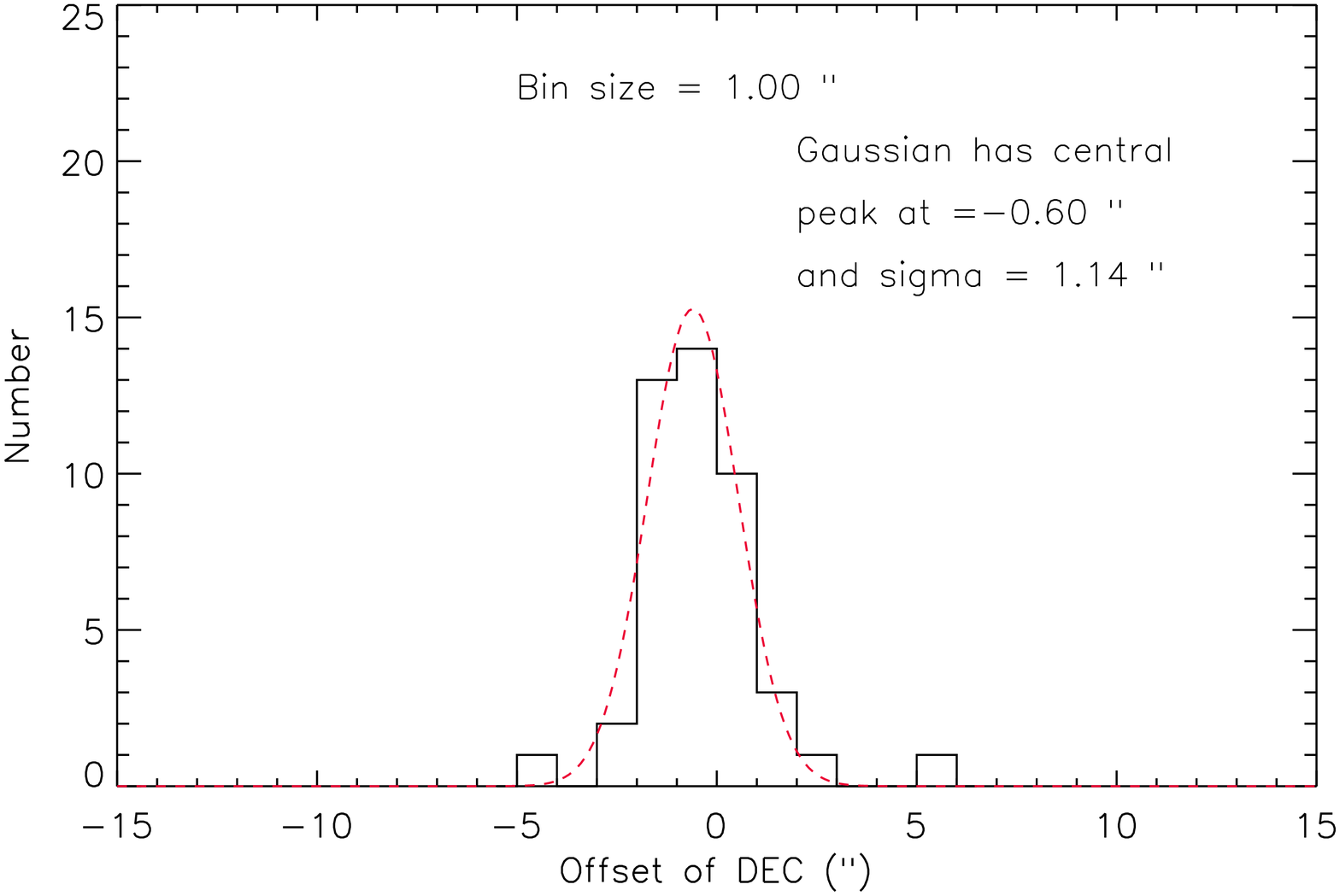}
\includegraphics[width=80mm, height=60mm]{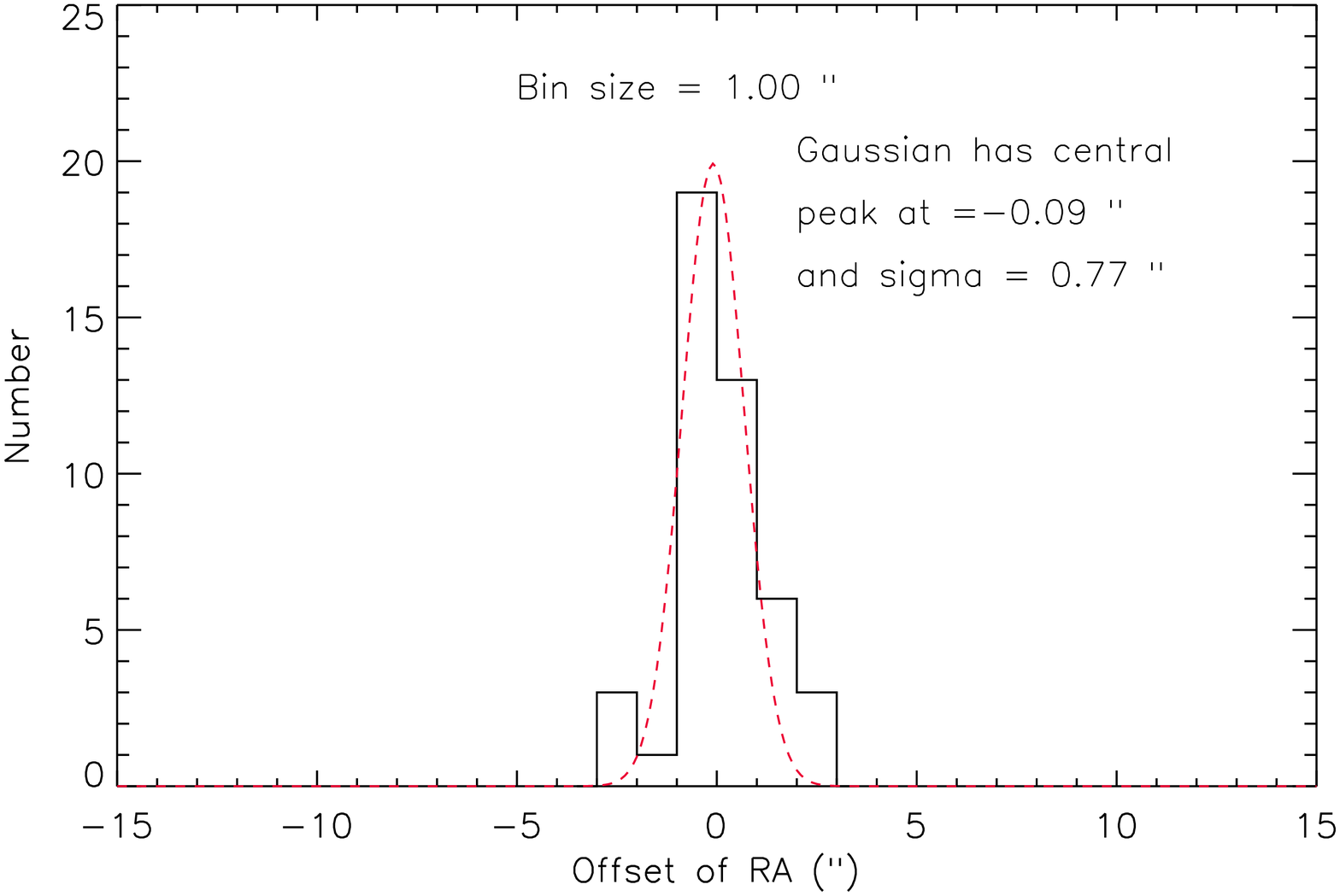}
\includegraphics[width=80mm, height=60mm]{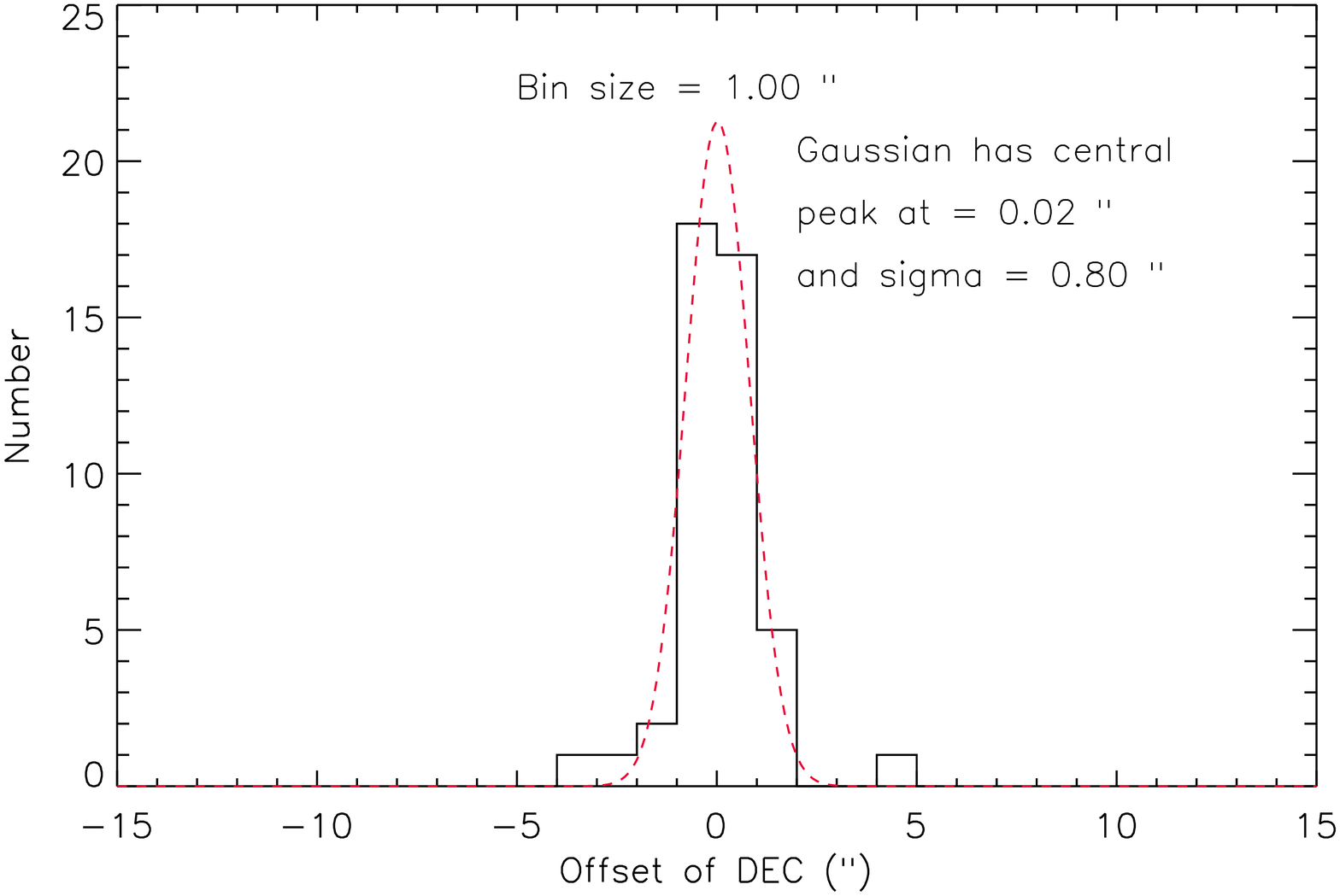}
\caption{Offset of TIMMI2 astrometry from MSX (top) and 2MASS (bottom) coordinates, in both RA and Dec, for the 45 objects in the 'clean' subsample that have astrometric information and clear 2MASS PSC counterparts. The dashed lines show gaussian fits to the data.}
\end{figure*}

\subsection{Photometry}

Aperture photometry was performed on all detected objects. For isolated point sources, an aperture of 2.4$^{\prime}$$^{\prime}$ was used, with sky annulus inner and outer radii of 3$^{\prime}$$^{\prime}$ and 4$^{\prime}$$^{\prime}$ respectively, whilst for extended sources and those near other sources, these were set individually. In the case of a G326.7249+00.6159 sources 1 $\&$ 2 and G298.1829$-$00.7860 sources 1 $\&$ 2, which have particularly small separations, aperture photometry was performed for the easiest source to separate and for both sources combined, with the flux for the second source obtained by subtracting the flux for the first source from the total flux. The errors in these fluxes were set based on the sky flux in the annulus of the smaller source, rather than the background levels in order to better reflect the uncertainty in these measurements. The standard observed nearest in time to a given object was used for calibration.

The estimated error in the flux measurements was obtained by repeating the measurement procedure with the standard aperture in 4 different background regions and calculating the average count levels. For frames with large amounts of extended emission, the error is likely to be larger as it is sometimes difficult to find areas free from source emission. In the case of non-detections, we have included a 3$\sigma$ upper limit based on the error calculated in this way. Comparison of standard star observations indicate that the overall photometric accuracy of our observations is 0.05Jy.

Figure 5 shows the comparison between MSX 8$\mu$m (left) and 12$\mu$m (right) fluxes and the derived TIMMI2 10.4$\mu$m flux for all detections, excluding the 18 targets which are MSX 12$\mu$m non-detections. The distribution of points in the comparison with the MSX 12$\mu$m band appears slightly squashed at lower fluxes, but this is is due to the fact that this band has a sensitivity limit of $\sim$1Jy, compared to the $\sim$0.1Jy limit of the MSX 8$\mu$m band. The solid line is a least squares bisector fit to the data and the dashed lines indicate perfect correlation (i.e. y = x) in order to guide the eye. Therefore, the difference between the two lines shows the difference between the MSX and TIMMI2 fluxes. We do not expect a perfect correlation between either MSX band fluxes and those obtained from our observations due to the larger beam size of the MSX observations. This results in extended emission that is not within our typical 2.4$^{\prime}$$^{\prime}$ aperture being included within the MSX flux, while it is minimised in the TIMMI2 photometry. Even for point sources we do not expect perfect agreement between MSX and TIMMI2 fluxes due to the differences in filter profiles and the rising slope of the spectral energy distribution towards longer wavelengths. The TIMMI2 filter includes the often deep 9.7$\mu$m absorption feature, and the MSX 8$\mu$m band contains some polycyclic-aromatic hydrocarbon (PAH) emission features, which GLIMPSE images (with far better spatial resolution than MSX at a similar wavelength) of similar regions show to be quite diffuse and extended \citep[e.g. M17][]{b30}. Nevertheless, there is a reasonably good correspondence between the MSX and TIMMI2 fluxes, with the spread being larger at lower fluxes due to the fact that diffuse emission in the MSX images becomes more of an issue for fainter objects.

\begin{figure*}
\centering
\includegraphics[width=80mm, height=60mm]{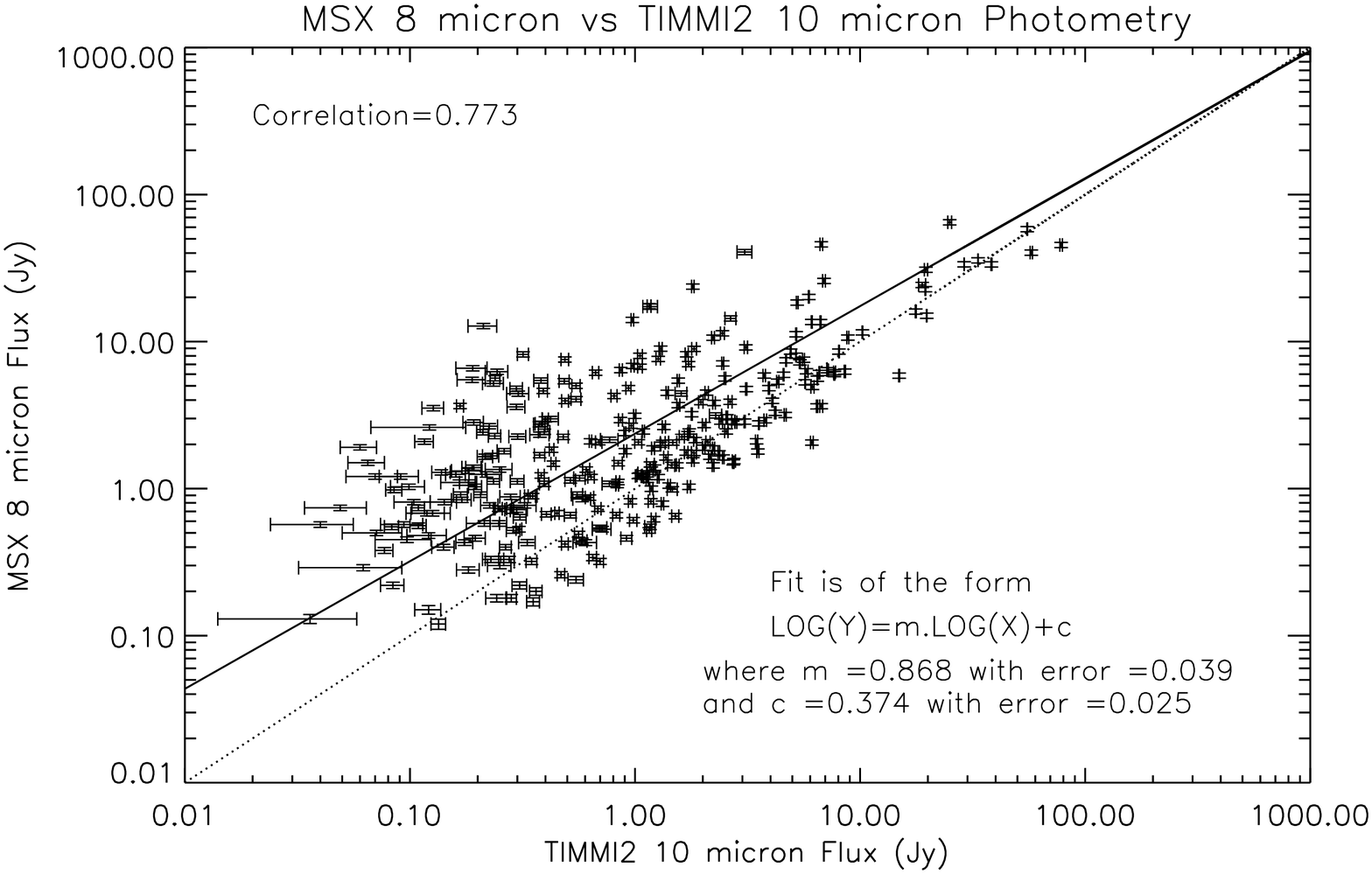}
\includegraphics[width=80mm, height=60mm]{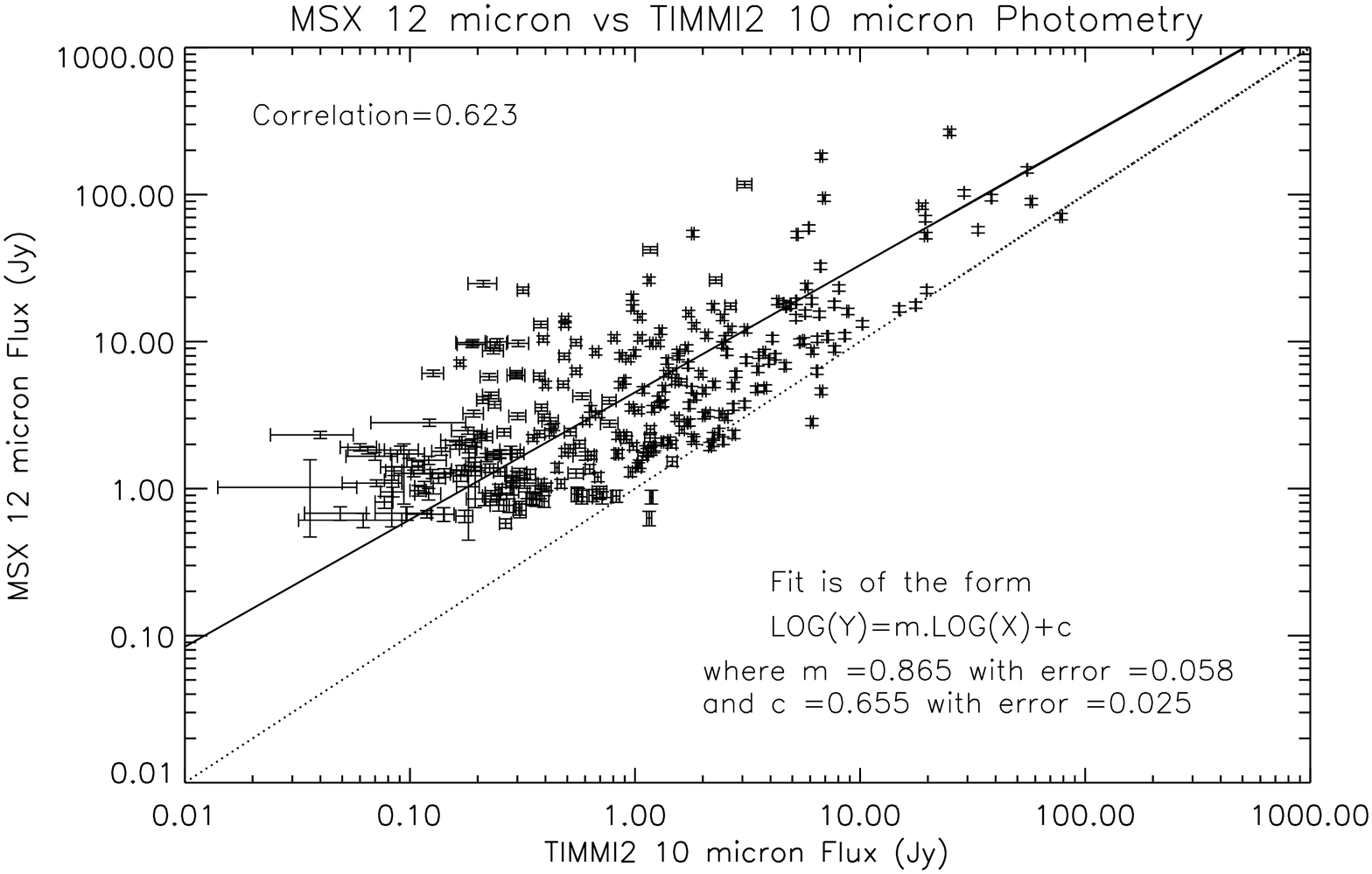}
\caption{Comparison of fluxes obtained from TIMMI2 observations with those from the MSX PSC for 8$\mu$m (left) and 12$\mu$m (right) for all targets with 12$\mu$m detections. The solid line corresponds to a linear bisector least squares fit to the data in log-log space, while the dashed line is the line y~=~x in order to guide the eye.}
\end{figure*}

In order to examine the effect of extended emission on the correlation between MSX and TIMMI2 fluxes, we again examine the 'clean' subsample mentioned previously ($\S$3.1), shown in figure 6. The solid line is a least squares bisector fit to the data and the dashed lines indicate perfect correlation (i.e. y = x) in order to guide the eye. From the better agreement between the fit to the data and the line of concordance, particularly for the comparison with the MSX 8$\mu$m data, it is clear that the fluxes more closely agree for this subsample than for all observations. This is to be expected, since these sources are mostly evolved stars and other isolated sources with little diffuse emission. The remaining difference reflects the filter profile and spectral differences already mentioned. The cut-off in the lower MSX 12$\mu$m flux is again due to the fact that the detector is less sensitive in this band than in the MSX 8$\mu$m band.

\begin{figure*}
\centering
\includegraphics[width=80mm, height=60mm]{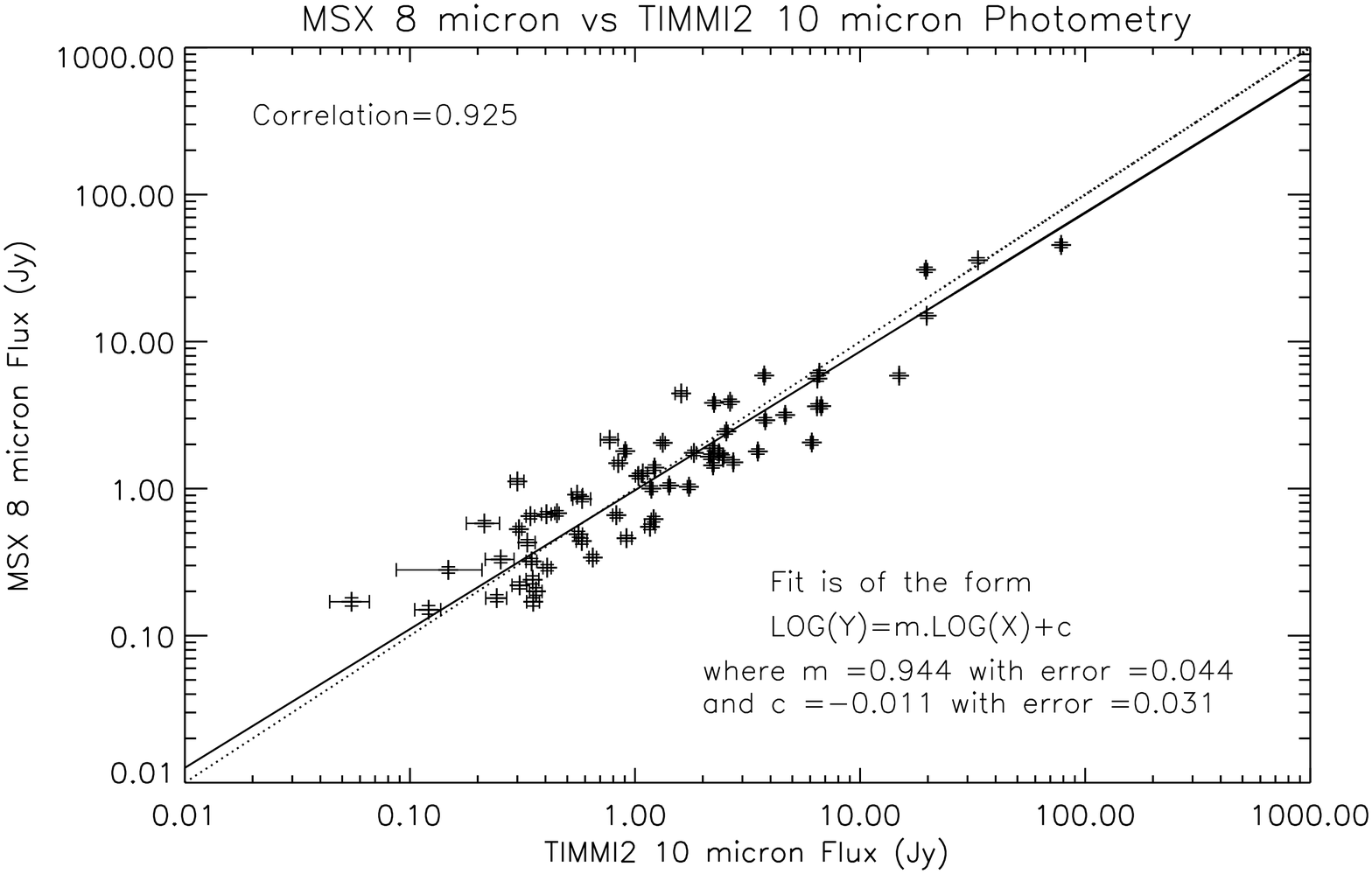}
\includegraphics[width=80mm, height=60mm]{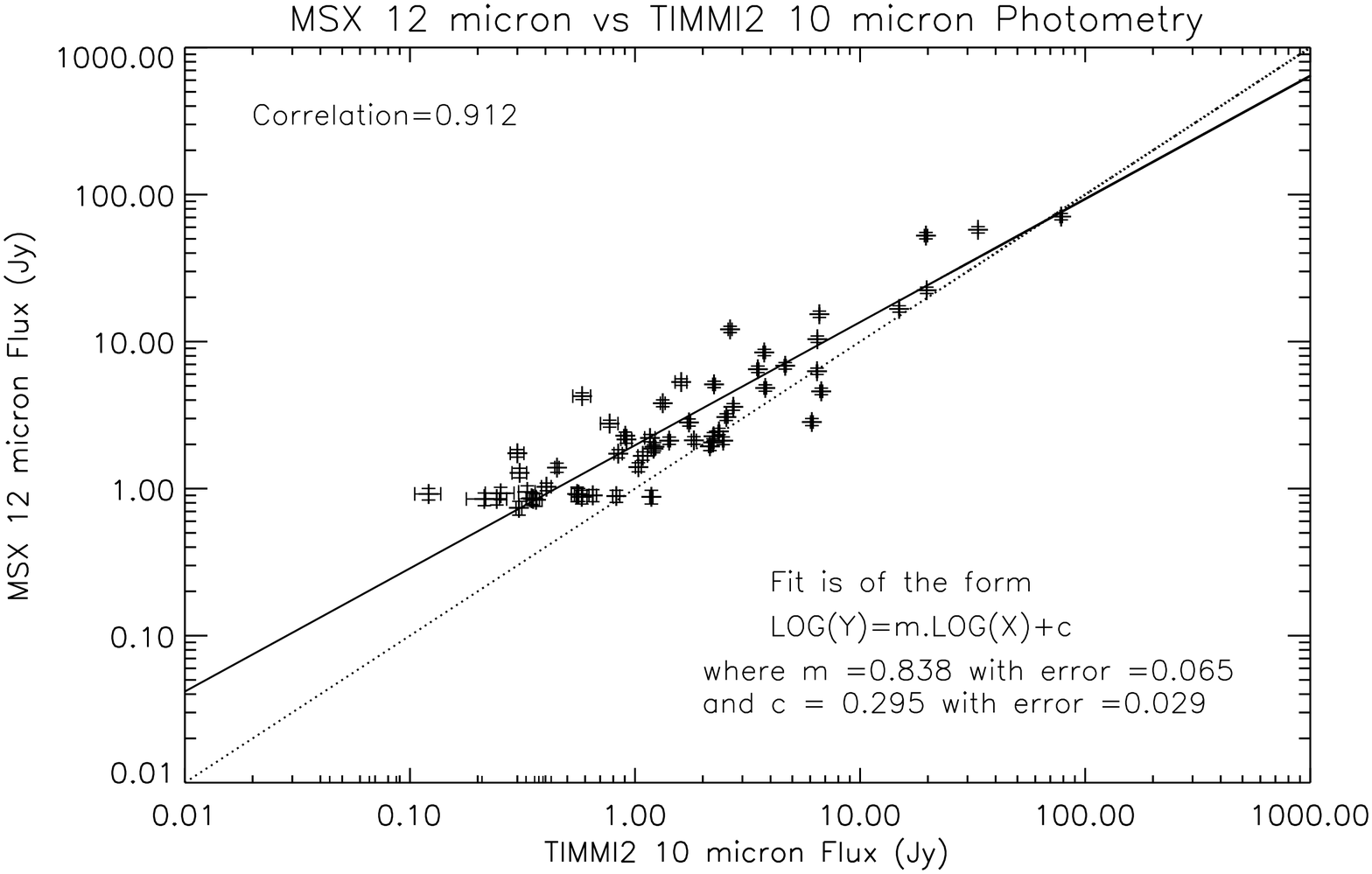}
\caption{Comparison of fluxes obtained from TIMMI2 observations with those from the MSX PSC for the 57 targets in the 'clean' subsample with MSX 12$\mu$m detections. The solid line corresponds to a linear bisector least squares fit to the data in log-log space, while the dashed line is the line y~=~x in order to guide the eye.}
\end{figure*}

\begin{table}
\centering
\caption{Sources also observed by \citet{b24}.}
\begin{tabular}{@{}cc@{}}
\hline
\centering
Our Object&\citet{b24}\\
Name&Source Number\\
\hline
G259.9395$-$00.0419&6,7\\
G268.4222$-$00.8490&11,12\\
G269.4582$-$01.4713&14\\
G291.2725$-$00.7198&16\\
G293.8282$-$00.7445&19\\
G300.9674+01.1499&21,22,23\\
G313.4573+00.1934&36\\
G345.0061+01.7944&53\\
\hline
\end{tabular}
\end{table}

\citet{b24} performed broad N (8-13$\mu$m) and Q (17.3-22.7$\mu$m) band observations of 8 of our targets (see table 3). They observe higher fluxes (on average approximately twice) what we do, due to the broader N band filter chosen for their observations (as opposed to our 9.46-11.21$\mu$m filter) which includes the PAH emission features at 8.7$\mu$m, more of the PAH feature at 11.3$\mu$m and is less dominated by the 9.7$\mu$m silicate absorption feature. Our astrometric positions agree with theirs to within their quoted error using guide stars of 7$^{\prime}$$^{\prime}$.

\subsection{Comparison with Radio Continuum Observations}

We compare our mid-IR imaging with the high resolution (1.5$^{\prime}$$^{\prime}$) radio continuum observations of \citet{b13} and \citet{b14}. 83 (24$\%$) have associated radio detections. In general, the agreement in alignment between our observations and the radio data is quite good, with some interesting cases where resolved radio continuum emission seems to trace an extended HII region while the mid-IR emission reveals some additional unresolved sources (e.g. G298.1829$-$00.7860 and G329.4211$-$00.1631). While one always has to be careful of projection effects, some of these may well be instances where the expanding HII region has triggered further star formation. The radio contours also show us that there are a few unresolved UCHII regions (e.g. G331.3546+01.0638) and PN (e.g. G292.7329$-$00.2482) within our targets, but that these only account for $\sim$7$\%$ and $\sim$3$\%$ of all unresolved mid-IR sources respectively. This justifies the main presumption that point sources in star formation regions are primarily MYSOs.

Within the ionised zone of a UCHII region, the gas is ionised by Lyman ($\lambda>$912nm) continuum photons from the central star and emits thermal bremsstrahlung in the radio. Resonantly scattered Lyman $\alpha$ photons within the ionised zone provide the main source of heating for the dust within this region, which re-emits the energy at infrared wavelengths. The radio and infrared emission should therefore be correlated.

In order to investigate whether any correlation between radio and mid-IR emission exists, we selected another subsample of sources. Starting from all sources where the radio and mid-IR sources seem to directly correspond, we eliminated objects which were substantially resolved ($\geq$8$^{\prime}$$^{\prime}$) in either the radio or mid-IR since the  6cm and 10.4$\mu$m fluxes are likely to be affected by either being resolved out or due to nod/chop effects. We also eliminate sources which are optically thick at radio wavelengths from the spectral index between our 3cm and 6cm observations, as the radio flux will not reflect the total nebular gas emission. Due to the criteria that the object be optically thin, so that we are observing all of the emission in the radio, we also eliminated those objects where we do not have both 3cm and 6cm observations. In addition, sources which are identified as PN within the RMS survey due to weak $^{13}$CO detections \citep{b34}, and/or their isolation and lack of extended emission in MSX and/or 2MASS images were eliminated from the subsample. This resulted in a subsample of 23 objects. The 6cm and 10.4$\mu$m fluxes for these sources are shown in figure 7, where we compare with fluxes taken from the calculated spectra for dusty HII region models of \citet[][their Fig. 8]{b15}. Overall our results agree well with the range of fluxes expected for HII regions based on these models, and are also consistent with the older simpler models of \citet{b40}. The two points which do not seem to follow the observed trend are G268.6162$-$00.7389 (radio quiet) and G348.6972$-$01.0263 (radio loud). 

\begin{figure*}
\centering
\includegraphics[width=80mm, height=60mm]{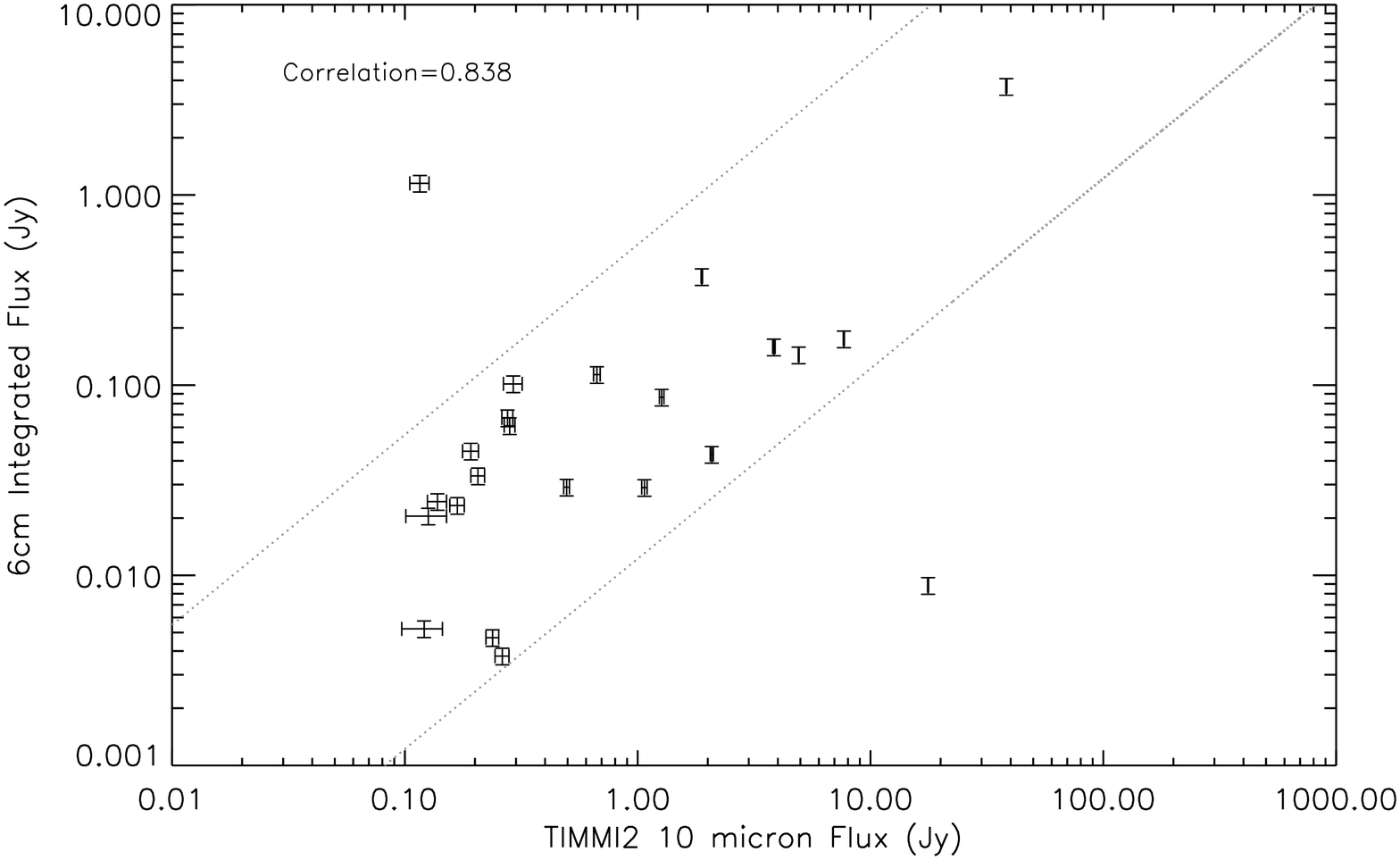}
\caption{10$\mu$m vs 6cm integrated radio flux for a subsample of sources. The dotted lines correspond to the minimum and maximum F$_{6cm}$/F$_{10\mu m}$ ratio from the calculated spectra for the dusty HII models of \citet[][their Fig. 8]{b15}. The majority of points lie within or near the expected region.}
\end{figure*}

G268.6162$-$00.7389 is associated with a stellar group \citep{b16} and with the reflection nebula Bran 225. The near-IR spectra of \citet{b36} show HeI, FeII and OI emission lines in addition to hydrogen recombination lines but the line ratios are too low to be case B. The $^{12}$CO(J$=$1$-$0) distance measurements of \citet{b23} suggest that it is at a distance of 2kpc, resulting in L$_{FIR}$=6682L$_{\odot}$ from IRAS photometry. All this data supports the classification by \citet{b36} of this source as a Herbig Be star, in which case the radio emission could be from a stellar wind. This is not in agreement with the optically thin spectral index of 0.12$\pm$0.01 returned from the radio observations of \citet{b13} which are more suggestive of a weak B type star HII region. However, the near-IR spectra are not consistent with that of a HII region. G348.6972$-$01.0263 lies within part of RCW122, and is definitely radio loud, since \citet{b14} detect even stronger radio emission than we do. \citet{b25} calculate a silicate optical depth at of $\tau$$_{9.7\mu m}$~=~3.5 for this region and \citet{b26} calculate A$_{V}$=117 and $\tau$$_{9.7\mu m}$$\sim$6.5 for the whole cloud, so there is certainly significant absorption towards this source. This suggests that the silicate absorption feature, as well as general extinction may play a part in the lower than expected 10.4$\mu$m flux of this source.

\section{Conclusions}

We have presented 10.4$\mu$m imaging observations for 346 candidate MYSOs in the southern hemisphere, primarily outside the GLIMPSE survey region. We obtain photometric and astrometric accuracies of $\sim$0.05Jy and 0.8$^{\prime}$$^{\prime}$ respectively. $\sim$64$\%$ of our observations contain unresolved sources, which are most likely either MYSOs or evolved stars. 24$\%$ of our observations contain only sources of extended 10.4$\mu$m emission (FWHM$>$1.6$^{\prime}$$^{\prime}$), which we expect primarily to be UCHII regions. For a subsample of our sources with radio continuum data we examine the F$_{6cm}$ - F$_{10\mu m}$ relation, with our results being consistent with theoretical predictions for UCHII regions. Our accurate astrometric data will aid near-IR spectroscopy in establishing object identities, and make clear which near-IR source is the counterpart of that seen at mid and far-IR wavelengths. A small fraction of point sources ($\sim$7$\%$) are unresolved UCHII regions, but these are readily identified by comparison with our radio continuum observations. In cases where multiple sources of 10.4$\mu$m emission are detected (25$\%$), our photometric data allow us to apportion the flux of larger beam observations between the different components, and so more accurately determine source luminosities.

Mid-IR observations of all the northern RMS sources not within the GLIMPSE region have been completed using Michelle on UKIRT and Gemini-North, and will be the subject of a future paper. Preliminary identification of all RMS sources has been completed and is available on the database. Near-IR spectroscopic confirmation is ongoing, as is resolving the remaining near/far and multi-component $^{13}$CO kinematic distances. Work has also begun on obtaining far-IR luminosity information for RMS sources without good quality IRAS PSC fluxes. We are currently exploring using the IRIS and IGA data products as well as MIPSGAL where available. This will allow us to finally assign luminosities to all confirmed YSOs, allowing a final catalogue to be published.

\begin{acknowledgements}

We would like to thank the anonymous referee for many helpful comments and suggestions which improved the clarity and content of this paper. We also thank Gaspare Lo Curto, Ivo Saviane and the rest of the staff at ESO's La Silla facility for their help over the course of our observations, particularly with the reference star pointing astrometry mode. JCM is funded by the Particle Physics and Astronomy Research Council of the United Kingdom (PPARC). This publication makes use of data products from the Two Micron All Sky Survey, which is a joint project of the University of Massachusetts and the Infrared Processing and Analysis Center/California Institute of Technology, funded by the National Aeronautics and Space Administration and the National Science Foundation.

\end{acknowledgements}

\label{lastpage}

\end{document}